\documentclass[fleqn,usenatbib]{mnras}

\usepackage{lineno}
\usepackage{xcolor}
\usepackage{newtxtext,newtxmath}

\usepackage[T1]{fontenc}

\DeclareRobustCommand{\VAN}[3]{#2}
\let\VANthebibliography\thebibliography
\def\thebibliography{\DeclareRobustCommand{\VAN}[3]{##3}\VANthebibliography}

\usepackage{graphicx}	
\usepackage[normalem]{ulem}
\usepackage{amsmath}	
\usepackage{multirow}
\usepackage{subcaption}
\usepackage{comment}
\usepackage{orcidlink}

\title[A New GW Probe to the Nature of Dark Energy]{A New Gravitational Wave Probe to the Nature of Dark Energy from the Aging of the Universe}
\author[Mukherjee]{
Suvodip Mukherjee\orcidlink{0000-0002-3373-5236}$^{1}$\thanks{suvodip@tifr.res.in}
\\
$^{1}$ Department of Astronomy \& Astrophysics, Tata Institute of Fundamental Research, 1, Homi Bhabha Road, Colaba, Mumbai 400005, India\\
}

\date{\today}
\pubyear{\the\year{}}

\begin{document}
	\label{firstpage}
	\pagerange{\pageref{firstpage}--\pageref{lastpage}}
	\maketitle
	
\begin{abstract}
One of the most dominant energy budgets in the Universe is Dark Energy, which remains enigmatic since its existence was first claimed based on observations of late-time cosmic acceleration. We propose a new way of inferring the dark energy equation of state (EoS) by measuring the aging of the Universe using only gravitational wave (GW) signals from coalescing binary compact objects of any masses. We show that the behavior of dark energy as the Universe ages will lead to a change in the observed chirp mass of GW sources inferred from observations of different stages of their coalescence. This change can
be studied by monitoring a coherent source over a few years, with two well-separated GW frequencies. With a coordinated network of GW detectors that can reach a sensitivity of Big Bang Observer, we can reach a $5\sigma$ detection of the dark energy EoS parameter $w_0=-1$ and its variation with cosmic time by using stellar origin binary black holes and binary neutron stars up to high redshift over 10 years of observation time without using any external calibrator. If the next generation of GW detectors can achieve this precision, then it can open a new window to discover the fundamental nature of dark energy.
\end{abstract}


\section{Introduction}\label{sec1}
The evidence of an accelerating Universe detected from cosmological observations has indicated the presence of about $75\%$ of the energy budget in dark energy \citep{1997ApJ...483..565P, 1998ApJ...507...46S, 1998AJ....116.1009R, 2003ApJS..148..175S, SDSS:2003eyi, Planck:2013pxb}. The nature of dark energy  {has remained unknown} for several decades, and multiple experiments are ongoing to understand it. Recently, the observation of Baryon Acoustic Oscillation (BAO) from DESI in combination with supernovae and CMB has hinted towards a redshift-dependent model of dark energy density, which can have a profound implication  {if the results are confirmed \citep{2024arXiv240403002D}}. We show in this paper that by using only  {gravitational-wave (GW)} signals from coalescing binary compact objects, we can reach an unprecedented precision to measure the nature of dark energy without requiring any calibration or  {astrophysical
assumption about the source population.} 

GWs-- a new cosmic messenger-- can probe the Universe up to high redshift from the current generation (such as Advanced-LIGO (Hanford, Livingston, and Aundha) \citep{KAGRA:2013rdx, LIGOScientific:2014pky}, Virgo \citep{VIRGO:2014yos}, KAGRA \citep{KAGRA:2020tym}) and next generation GW detectors (such as Cosmic Explorer \citep{Hall:2020dps}, Einstein Telescope\citep{2010CQGra..27h4007P, Maggiore:2019uih}, LISA \citep{2017arXiv170200786A}, and concepts such as LGWA \citep{Ajith:2024mie}). As the signal emitted from coalescing binary compact objects is extremely well modeled in the General Theory of Relativity (GR), it makes it possible to accurately calculate the change in the GW frequency and amplitude with time  {as a function of the source properties}. As a result, the GW sources are both standard clocks and standard sirens, which make it possible to accurately measure the chirp mass and luminosity distance of the GW sources \citep{1986NaturS, PhysRevD.57.4535}.

In this work,  {we show that we can measure the nature of
dark energy without requiring any additional information by using two
salient properties: } (i) we can measure both redshifted chirp-mass and luminosity distance to the GW sources with reasonable accuracy and (ii) the period of observation of the GW signals from a single source can be measured in phase for a year(s) with the aid of multi-band GW observatories operating at different GW frequencies. 

\section{Cosmology from the aging of the Universe}\label{sec2}
The measurement of the expansion rate of the Universe at different cosmic epochs denoted by $H(z) \equiv \dot a(t)/a(t)$, where $a(t)/a(t_0)\equiv 1/(1+z)$\footnote{$t_0$ denotes the  present time of the Universe.} is the scale factor at a redshift $z$,  {enables} a profound understanding of the constituents of the Universe through the relation
\begin{equation}\label{eq1}
   H(z) = H_0\bigg[\Omega_{\rm dm}(z) + \Omega_{\rm b}(z) + \Omega_{\rm de}(z) + \Omega_{\rm rad}(z) \bigg]^{1/2},
\end{equation}
where $H_0$ is the Hubble constant which captures the expansion rate today in the Universe, and $\Omega_i(z)\equiv \rho(z)/\rho_c$ denotes the energy density in the $i^{th}$ constituent of the Universe with respect to the critical energy density $\rho^2_c \equiv 3H_0^2/8\pi G$. As per our current understanding, about $95\%$ of today's Universe is made up of dark energy (denoted by $\Omega_{\rm de}$) and cold dark matter (denoted by $\Omega_{\rm dm}$), which are still unknown to us, and the remaining $5\%$ contribution comes from baryons (denoted by $\Omega_{\rm b}$) and radiation (denoted by $\Omega_{\rm rad}$) \citep{Planck:2018vyg}. 
\begin{figure}
     \centering
\includegraphics[width=0.5\textwidth]{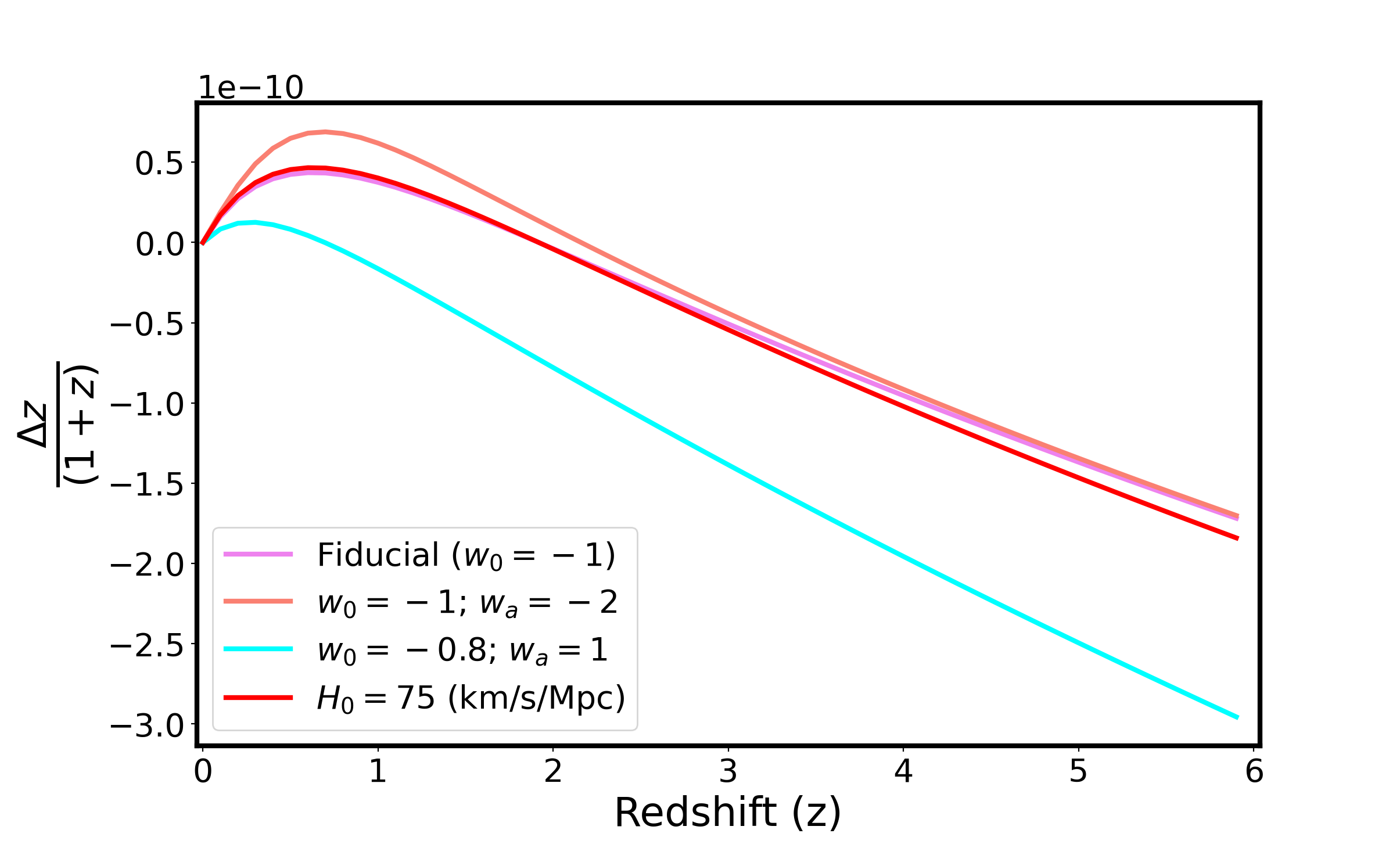}
    \caption{The relative variation of redshift due to different models of cosmic expansion history is shown between two epochs separated by $\Delta t=5$ years  {using Eq. \eqref{eq:zdef}.} The fiducial cosmological model is considered for the LCDM model with Planck-2018 cosmological parameters and $H_0= 70$ km/s/Mpc.}
    \label{fig:cosmo-theory}
\end{figure}
 {However, these fractions change at earlier epochs of the Universe, and measuring these fractions at different cosmic epochs
would enable us to better understand the nature of these quantities -- particularly dark matter and dark energy, which are currently not well understood.} The time-variation of dark energy is usually modeled by a vanilla parametric form of the dark energy EoS $w(a)= w_0 + w_a (1-a(t))$ 
where $w_0$ is a constant and $w_a$ captures the time evolution of the dark energy  {equation of state} (EoS) in terms of the cosmological scale factor $a(t)$ \citep{Chevallier:2000qy, PhysRevLett.90.091301}. The  {current} best-fit cosmological model supports $w_0=-1; w_a=0$ (cosmological constant) \citep{Planck:2018vyg}. However, the recent DESI observations have indicated the signature of  {a redshift dependent dark energy} showing a best-fit value of $w_0>-1; w_a<0$ \citep{2024arXiv240403002D}. Furthermore, the limitation of this vanilla parametric form should not be neglected \citep{Shlivko:2024llw}. 


Direct evidence of cosmic expansion can also be inferred from the change in the aging of the Universe. This effect  {reflects a change in the cosmic redshift with time. Given the relation between} redshift and scale factor $z= a(t_0)/a(t) -1$, we can write the change in the redshift of a source with respect to the observed time $t$ as
\begin{equation}\label{eq3}
    \frac{d\ln (1+z)}{dt}= H_0 -\frac{H(z)}{(1+z)}. 
\end{equation}
This implies that the fractional change in the redshift to the source depends on the difference between the Hubble parameter today (at $z=0$) and at a redshift $z$. An observation that can measure the change in the redshift of a source between an observation time separated by $\Delta t$ precisely, can probe the expansion history. In Fig. \ref{fig:cosmo-theory}, we show the variation in redshift for different cosmological models for $\Delta t=5$ years. This effect on galaxies was previously calculated by \citep{1962ApJ...136..319S, 1998ApJ...499L.111L}. 

\section{Gravitational wave as a probe to cosmic aging}\label{sec3}
 {The} GW signal in terms of frequency $f$ from a coalescing binary of masses $m_1$ and $m_2$ situated at a luminosity distance $d_L$ can be written in the inspiral phase in terms of the source-frame chirp mass $\mathcal{M}= (m_1m_2)^{3/5}/(m_1+m_2)^{1/5}$ as \citep{1989thyg.book.....H, Poisson:1995ef}
 \begin{equation}\label{strain}
   h_{+,\times}(f, \hat n)= \sqrt{\frac{5}{96}}\frac{G^{5/6}\mathcal{M}^2 (f\mathcal{M})^{-7/6}}{c^{3/2}\pi^{2/3}d_L}{\Theta}_{+, \times} (\hat L.\hat n),
 \end{equation}
 where ${\Theta}_{+,\times} (\hat L.\hat n)$ is the factor that captures the projection between the orbital angular momentum $\hat L$ and the direction to the line of sight to the observer $\hat n$ for the plus ($h_+$) and cross-polarization ($h_\times$) states of the GW signal. A key property of coalescing binary objects is the change in the frequency of the signal with time \citep{1989thyg.book.....H, Poisson:1995ef}
 \begin{equation}\label{chirp1}
     \frac{df}{dt}= \frac{96\pi^{8/3}}{5c^5}(G\mathcal{M})^{5/3}f^{11/3}.
 \end{equation}
This shows that the change in the frequency of the GW signal depends strongly on the chirp mass of the coalescing binaries and the instantaneous frequency of the GW signal. In an expanding Universe, the time gets dilated as $dt_o= dt(1+z)$ and frequency gets redshifted as $f_o= f/(1+z)$\footnote{ {Here $dt_o$ and $f_o$ denotes the observed time difference and observed frequency respectively.}}, which leads to the redshifted observed chirp mass $\mathcal{M}^{\rm det}= (1+z)\mathcal{M}$.  This implies 
 \begin{equation}\label{chirp2}
     \frac{df_o}{dt_o}(\tau)= \frac{96\pi^{8/3}}{5c^5}(G(1+z(\tau))\mathcal{M})^{5/3}f_o^{11/3},
 \end{equation}
 \begin{figure*}
    \centering
    \includegraphics[width=1.0\textwidth]{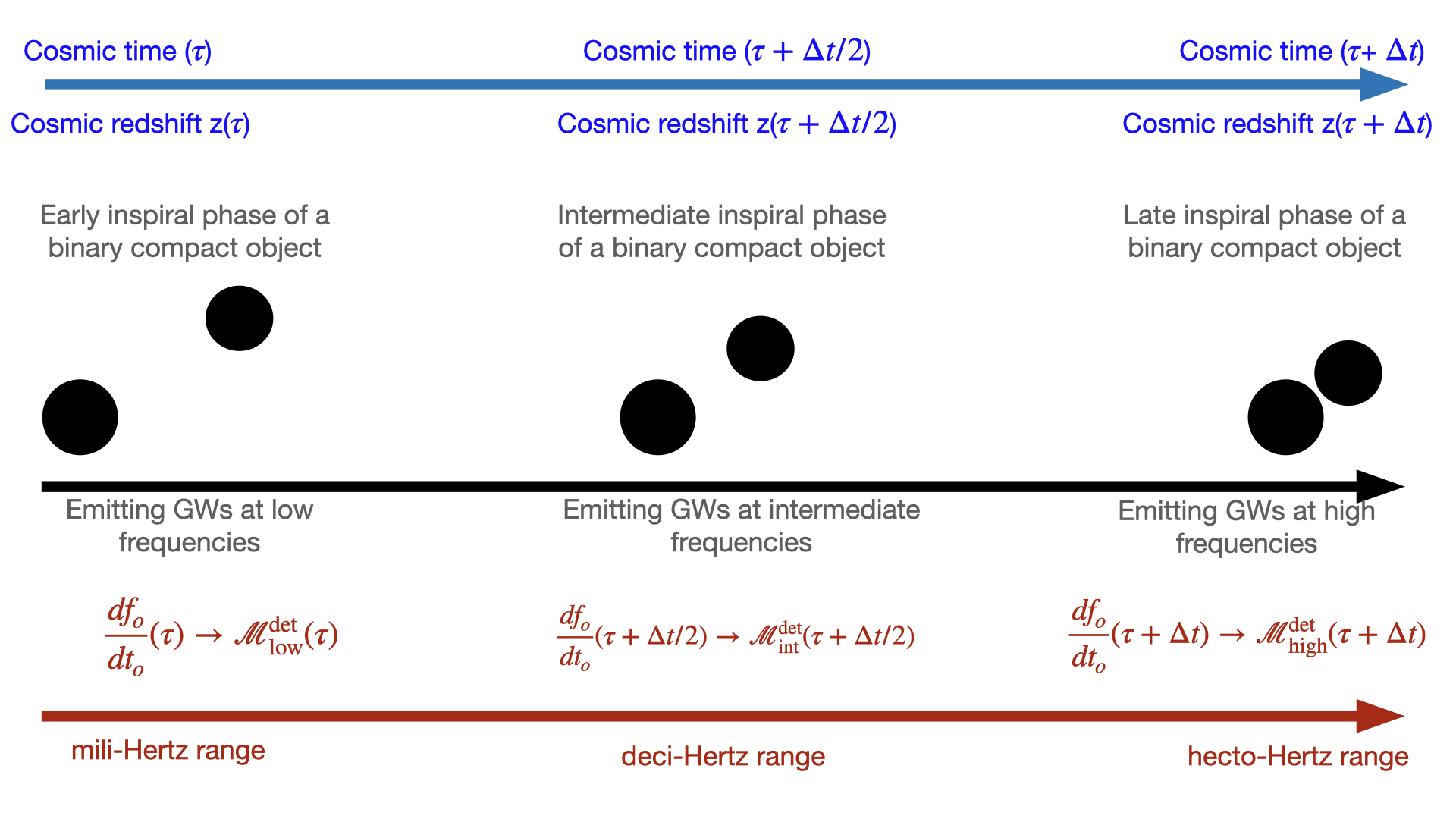}
    \caption{This schematic diagram explains the key principle behind this technique to measure the expansion history of the Universe. A GW source emitting at a low frequency witnesses a different cosmic expansion history than when it is emitting at a later time at a high frequency due to the aging of the Universe. As a result, the inferred detector-frame chirp mass from low frequency $\mathcal{M}^{\rm det}_{\rm low} (\tau)= (1+z(\tau))\mathcal{M}$ will be different from the detector-frame chirp mass from high frequency $\mathcal{M}^{\rm det}_{\rm high} (\tau + \Delta t)= (1+z(\tau + \Delta t))\mathcal{M}$ due to the aging of the Universe.}
    \label{fig:motivation}
\end{figure*}
where $\tau$ denotes a cosmic epoch.  {The above equation shows that, for a
given GW source, the rate of evolution of its GW frequency in the early
inspiral phase and late inspiral (or in the merger/ringdown) phase will
be happening at a different cosmic redshift.} As a result, the corresponding chirping behavior (Eq. \eqref{chirp2}) of the signal will be  {tracing a very slightly different value} of $z(\tau)$ due to the cosmic expansion (see the schematic diagram in Fig. \ref{fig:motivation}). 

So if we observe a coherent GW source detected at a low frequency during an early inspiral phase and also at a high frequency during the late inspiral phase, they will have a little different detector-frame chirp mass denoted by $\mathcal{M}^{\rm det}_{\rm low/high}(\tau)= (1+z(\tau))\mathcal{M}$. Their relative difference can be connected to cosmic expansion history as
\begin{equation}
    \begin{aligned}
    \Delta \mathcal{M}^{\rm det}_ {{\rm high,low}}\equiv&   \frac{\mathcal{M}^{\rm det}_{\rm high}(\tau+\Delta t)}{ \mathcal{M}^{\rm det}_{\rm low}(\tau)}-1 = \frac{z(\tau+\Delta t)-z(\tau)}{1+z(\tau)},\\ 
    \Delta \mathcal{M}^{\rm det}_ {{\rm high,low}} =&\Delta t \bigg(H_0 - \frac{H(z(\tau))}{1+z(\tau)}\bigg),
        \end{aligned}
\end{equation}
where,  in the last equation, we have used Eq. \eqref{eq3} to show its relation with the change in the cosmic expansion rate as
 {\begin{equation}\label{eq:zdef}
    \begin{aligned}
\frac{z(\tau+\Delta t)-z(\tau)}{1+z(\tau)}=\Delta t \bigg(H_0 - \frac{H(z(\tau))}{1+z(\tau)}\bigg).
        \end{aligned}
\end{equation}}
The value of $\Delta t$ for GW sources such as stellar origin BBHs can be a few years if a source is observed in their inspiral stage at frequencies $10^{-3}$ Hz and also at a later stage in their inspiral when they are emitting at $\sim 10^{2}$ Hz. Combining the relative chirp mass difference measurement with another well-measured quantity, the luminosity distance to the source that is inferred from the average of the signal from the complete observation period, we can infer the expansion history of the Universe and cosmological parameters without requiring any EM counterpart, small sky-localization error, additional galaxy-catalog, or making any assumption on the GW mass population. A few key advantages of this  method are: 
\begin{enumerate}
       \item Due to the coherent nature of the GW signal and its robust predictability from GR, the calibration of the measured luminosity distance and chirp mass is not required. 
    \item This method allows for the study of cosmology with only GW sources without requiring an independent redshift measurement or better sky localization of the sources. 
    \item The method can robustly probe expansion rate up to a very high redshift without any assumption on the astrophysical population of sources. 
\end{enumerate}

\begin{figure}
    \centering
\includegraphics[width=0.5\textwidth]{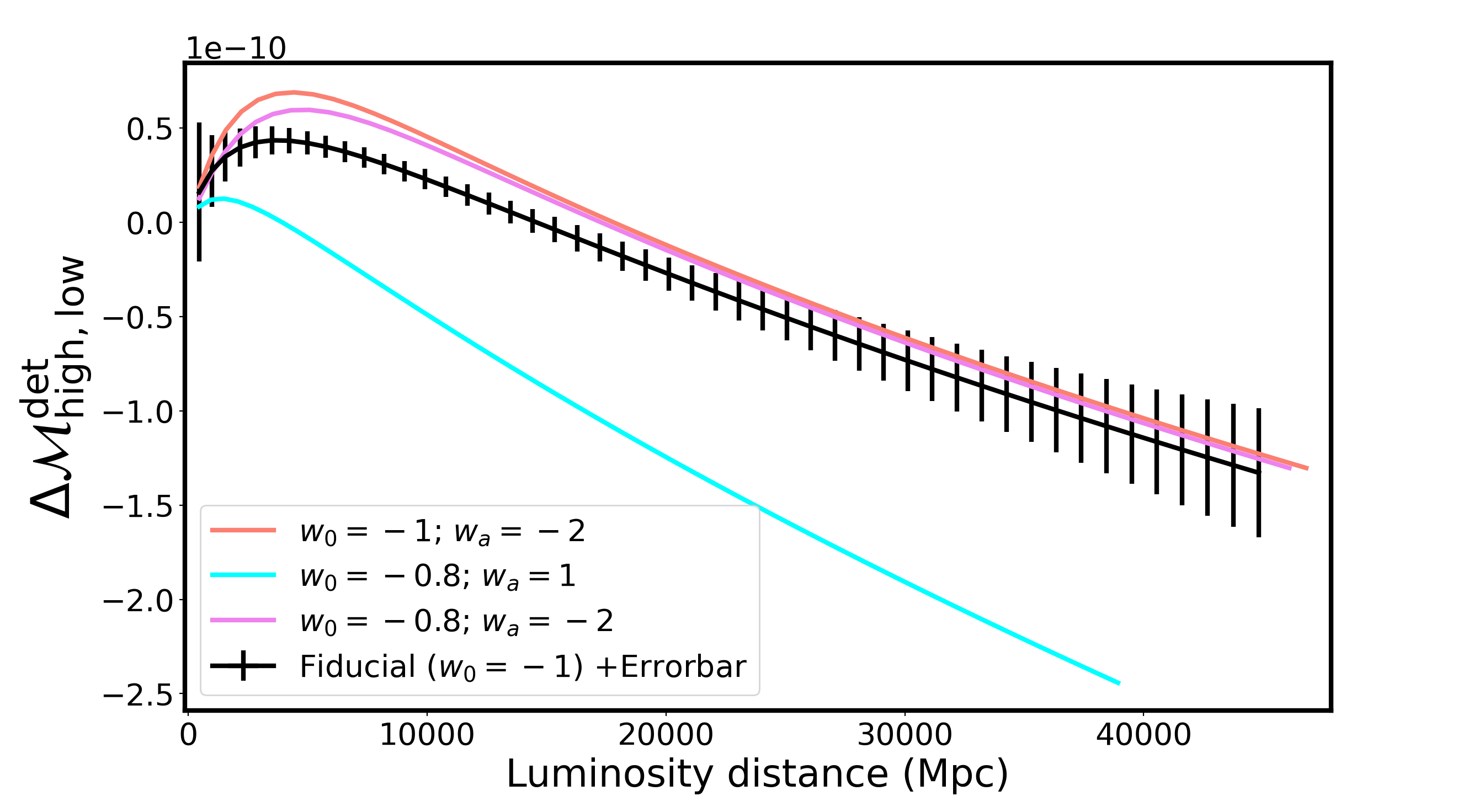}
 \caption{We show the difference in the measured
detector-frame chirp mass, inferred from observations of GW sources
in GW frequencies ~$10^{-2}$ Hz (low) and ~$10^{2}$ Hz (high) which
are separated by a time difference $\Delta t = 5$years. We have
assumed BBH observations integrated over a period of 10 years for a
chirp mass error of $10^{-9}$ and relative luminosity distance error of
$10^{-1}$. Results are shown for different cosmological models with
the measurement uncertainties also shown for the fiducial LCDM
model.}\label{fig:m-dl}
\end{figure}
\section{Results}\label{sec4}
The measurement of the high redshift expansion rate of the Universe demonstrated in the last section is possible from coherent GW sources such as binary neutron stars (BNSs), neutron star-black holes (NSBHs), binary black holes (BBHs) of all masses, which can include black holes of both astrophysical and primordial origin. In this analysis, we only consider stellar origin BBHs, as it is a population that can be detected up to high redshift, and also we have an estimate of its merger rate from LIGO-Virgo-KAGRA collaboration observations \citep{KAGRA:2021duu}.  We obtain the number of GW mergers at any redshift by integrating the merger rate density $R(z)$ of the GW sources of some masses $m_1, m_2$ over redshift by multiplying over the comoving volume $dV/dz$ and operational time $T_{\rm obs}$ of GW detectors as 
\begin{equation}\label{eq:merg}
    N_{\rm GW}= T_{\rm obs}\int \frac{R(z)}{1+z}\frac{dV}{dz} dz.
\end{equation}
\begin{figure}
    \centering
    \includegraphics[width=0.4\textwidth]{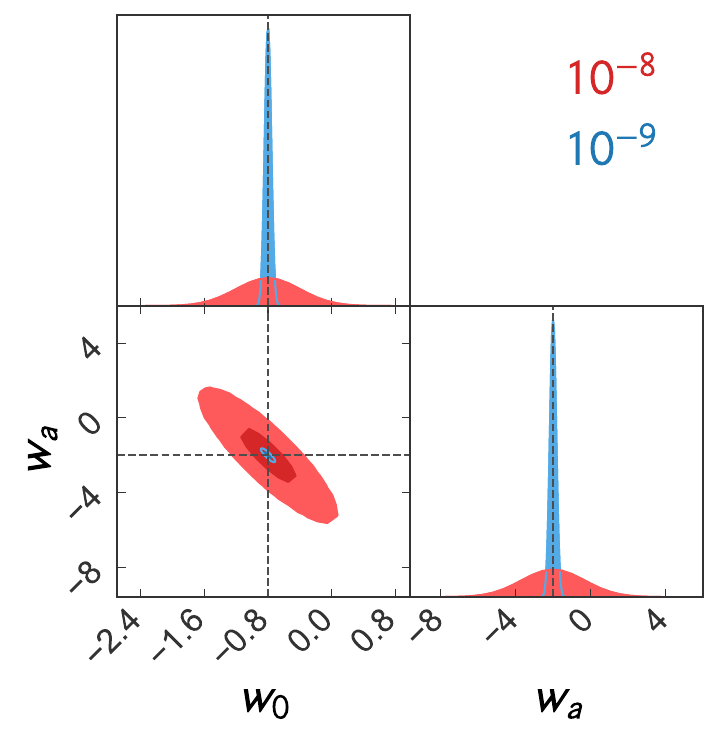}
    \includegraphics[width=0.4\textwidth]{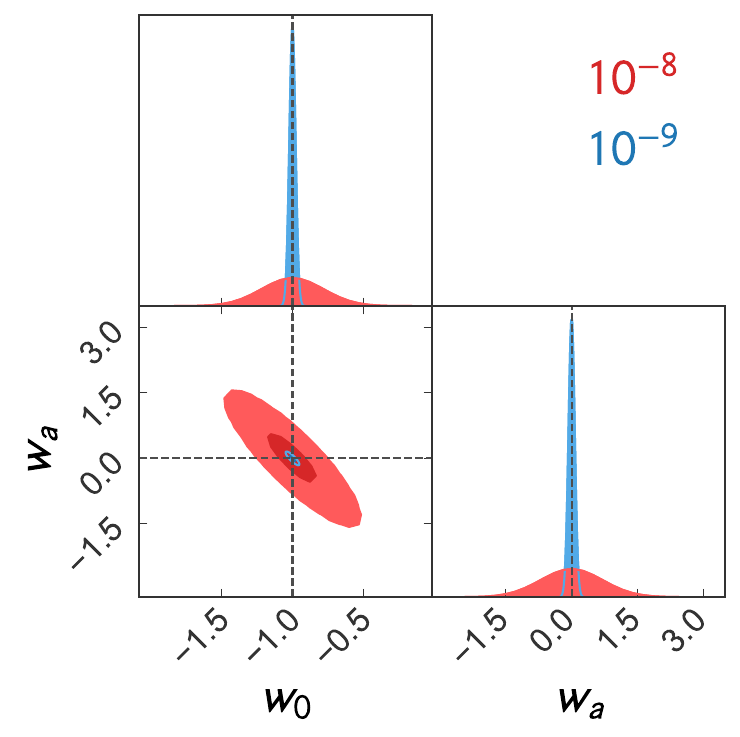}
    \caption{The figure shows the Cramer-Rao bound on the dark energy EoS $w_0=-0.8$ and $w_a=-2$ (left) and $w_0=-1.0$ and $w_a=0$ (right) for two different relative chirp mass errors (shown in red ($10^{-8}$) and blue ($10^{-9}$)), after combining 10 years of observation of stellar origin BBHs up to $z=5$.}
    \label{fig:de}
\end{figure}
We model the merger rate of stellar origin BBHs $R(z)$ assuming that the BBHs follow the Madau-Dickinson star formation rate \citep{Madau:2014bja}. The local merger rate $R(z=0)$ is taken as 20 $\rm{Gpc}^{-3}\, \rm{yr}^{-1}$ as per the third observation run of LVK Collaboration (GWTC-3) \citep{KAGRA:2021duu}. This gives us in total about $10^5$ BBHs per year across all redshifts. This number can increase by about a factor of 2-3 if we include the upper bound of BBH merger rate from GWTC-3 \citep{KAGRA:2021duu}.  {Similarly, for BNS sources, the number of detectable sources can be between $10^5$-- $10^7$ per year \citep{Ronchini:2022gwk, Dhani:2023ijt},   considering the uncertainty from GWTC-3\citep{KAGRA:2021duu}. However, it is important to note that these rates are based on theoretical estimates, and the actual rates of merger can be higher due to lack of understanding of the high redshift Universe and also non-astrophysical formation channels of compact objects, such as primordial black holes, for which the merger rates can be higher by orders of magnitude \citep{Raidal:2017mfl, Ali-Haimoud:2017rtz}. In this analysis, we only focus on astrophysical sources, with a pessimistic merger rate. But if the merger rate at high redshift is higher, the estimate will improve.}

The measurability of the expansion rate using this technique from a single GW event depends on the fractional standard deviation in measuring the masses in the low and high frequencies as 
\begin{equation}
    \Sigma_{\Delta_M}= \sqrt{\bigg(\frac{\sigma_\mathcal{M_{\rm low}}}{\mathcal{M_{\rm low}}}\bigg)^2+ \bigg(\frac{\sigma_\mathcal{M_{\rm high}}}{\mathcal{M_{\rm high}}}\bigg)^2},
\end{equation}
where $\frac{\sigma_\mathcal{M_{\rm low, high}}}{\mathcal{M_{\rm low, high}}}$ are the relative error on the mass measurements from the low frequency and high frequency of GW data. To explore the measurability of this signal, we consider two scenarios for the relative error on the chirp mass namely, $\frac{\sigma_\mathcal{M_{\rm low, high}}}{\mathcal{M_{\rm low, high}}}=10^{-8}$ and $10^{-9}$ for two frequency range of the GW inspiral stage which is separated by $\Delta t=5$ years. For comparison, the expected relative uncertainty on the chirp mass from current proposals  \citep{2010CQGra..27h4007P,2017arXiv170200786A,Hall:2020dps,Ajith:2024mie}, is of the order $10^{-5}- 10^{-6}$. 
So, the uncertainty in the chirp mass required for this analysis is about $10^2- 10^3$ better than the currently proposed experiments. Though this improvement in uncertainty is large, the feasibility of this with the advancement of GW detector technology can make groundbreaking discoveries, as will be discussed later in the paper. In particular, from the measurement perspective, precise inference of $\dot f/f$ over a broad range of frequencies is required for this measurement. Along with precise chirp mass inference, the precision on luminosity distance should be high as well. The relative luminosity distance uncertainty from individual GW sources ($\Sigma_{d_L} \equiv [(\sigma^{\rm det}_{d_L})^2+ (\sigma^{\rm wl}_{d_L})^2]^{0.5}/d_L$) will arise primarily from two sources, uncertainty due to detector noise $\sigma^{\rm det}_{d_L}$ and due to weak lensing $\sigma^{\rm wl}_{d_L}$. For a relative luminosity distance uncertainty of about $\sigma^{\rm det}_{d_L}/d_L= 1\%$, the dominant source of uncertainty is due to weak lensing for GW sources at high redshift ($z>1$) \citep{2010PhRvD..81l4046H}. The requirement on the luminosity distance uncertainty for this technique is not stringent in comparison to the expected distance uncertainty from the next-generation GW detectors \citep{2010CQGra..27h4007P,2017arXiv170200786A,Hall:2020dps,Ajith:2024mie}. 

We estimate the error in the inference of time-dependent chirp mass $\Delta \mathcal{M^{\rm det}_{\rm high, low}}$ as a function of luminosity distance in Fig. \ref{fig:m-dl} for the fiducial case of LCDM model ($w_0=-1$ and $w_a=0$) and $\frac{\sigma_\mathcal{M_{\rm low, high}}}{\mathcal{M_{\rm low, high}}}=10^{-9}$  {on individual GW sources}. The plot shows that by using only observations of chirp mass with luminosity distance, we can measure the expansion history for a typical value of the astrophysical population of BBHs using the technique proposed in this work. 
We further explore the prospect of this technique to measure the phenomenological parameters $w_0, w_a$, which capture the dark energy EoS and its redshift evolution. In Fig. \ref{fig:de} we show the Cram\'er-Rao bound \citep{1945r, 1946c} on the minimum error in the inference of the dark energy EoS parameters ( $w_0, w_a$) using Fisher analysis. The red and blue contours show the expected uncertainty for two cases of chirp mass errors $10^{-8}$ and $10^{-9}$ respectively, with 10 years of observation period for two choices of the dark energy EoS parameters, (a) $w_0= -0.8$; $w_a=-2$, and (b) $w_0= -1$; $w_a=0$. The plot indicates that with about $10^{-8}$ relative error on chirp mass, we can reach about a $5$-$\sigma$ detection of dark energy EoS within 10 years of the observation period. However, this measurement can reach a percent level ($\sim 2\%$) measurement of dark energy EoS and its redshift evolution for a relative chirp mass uncertainty $10^{-9}$. As a result, it will make a definite discovery on the nature of dark energy using observational probes that are free from any external calibration.  If the number of GW sources detectable per year in multiple GW bands is more/less than $10^5$, then the required uncertainty on chirp mass measurement for the same precision to measure dark energy EoS will scale as $10^{-9}\sqrt{N_{GW}/10^5}$. This method can also explore the Hubble constant and other cosmological parameters, but we restrict this analysis only to dark energy, as it is one of the unsolved problems in cosmology.

 {So far, we have particularly discussed the cosmological signal and the required instrument noise for making such measurements. However, it is also important to gauge whether there can be any astrophysical systematic that can contaminate the signal. One such source of astrophysical contamination can arise from the local acceleration of the binary toward the galaxy. One of the most dominant sources of such effect is the local acceleration due to the property of the host, such as whether it is in a globular cluster or in a massive galaxy having a supermassive black hole at the center. The contribution from local acceleration along the line of sight can lead to a relative redshift uncertainty as \citep{1987gady.book.....B, PhysRevD.96.063014}
\begin{equation}
    \delta z=  \frac{\vec a_{\rm source}.\hat n}{c}\Delta t,
\end{equation}
where, $\vec a_{\rm source}.\hat n$ denotes the acceleration along the direction of the observer denoted by $\hat n$, and $\Delta t$ denotes the time window over which the signal is observed. Assuming that the source is at a distance $s$ from the center and going through a velocity dispersion $v_s$, we can then write the above equation using $\hat a_{\rm source}.\vec n= \cos\theta$ as
\begin{equation}
\begin{split}
        \frac{\delta z}{1+z}&=  \frac{v_s^2\cos \theta}{c(1+z)}\frac{1}{s}\Delta t,\\
        &\sim 10^{-11}\cos \theta\bigg(\frac{v_s}{100 \rm km/s}\bigg)^2\bigg(\frac{1 \rm kpc}{s}\bigg)\bigg(\frac{1}{1+z}\bigg)\bigg(\frac{\Delta t}{\rm year}\bigg).
\end{split} 
\end{equation}
As GW sources will have a random orientation with respect to the observer, this term vanishes on averaging over many (about $10^5$) sources due to the dependence on the cosine of the angle between the observer and the direction of acceleration. Also, the associated contamination to the error is about 2-3 orders of magnitude smaller than the error on chirp mass $10^{-8}$--$10^{-9}$ considered in this analysis. Even if a few sources are very near the galactic center (within a few tens of pc), those sources will be outliers and won't contribute to the all-sky averaged uncertainty significantly. However, the GW waveform systematics needs to be well in control with a relative uncertainty better than $10^{-11}$ for measuring this signal.}

\section{Future prospect of measurement}
 {The science case proposed in this work provides a unique opportunity to study the nature of dark energy without any calibration. Such measurements will be only possible from GW sources at this precision if it can be achieved by the upcoming detectors. The key requirement of this measurement is a controlled uncertainty on the frequency measurement of the GW sources over different frequency bands of the GW signal. Such measurement will be possible from detectors that can be operational in the deci-hertz range with possible overlap with the hecto-hertz. Such sources can detect less than ten solar mass BBHs and BNSs from about a few tens of mili-Hz to about a few tens of Hz. As a result, the continuous monitoring of the sources will be possible.} 

 {Currently, we are at a stage of planning for the next-generation GW detectors and setup from both ground and space, covering from mili-hertz to kilo-hertz, such as 
Big Bang Observer (BBO) \citep{Crowder:2005nr}, Deci-hertz Interferometer Gravitational-Wave Observatory (DECIGO) \citep{Seto:2001qf,Kawamura:2006up}, advanced TianGO\citep{Kuns:2019upi}, Atomic Experiment for Dark Matter and Gravity Exploration in Space (AEDGE)\citep{Ellis:2020lxl}, Gravitational-wave Lunar Observatory for Cosmology (GLOC)\citep{Jani:2020gnz}, Lunar Gravitational-wave Antenna (LGWA)\citep{Ajith:2024mie}
. The new avenue to explore dark energy proposed in this work can be one of the key science cases for these detectors, with the appropriate design of the instrument which can meet the science requirement of $\sim \sigma_{\mathcal{M}}/\mathcal{M}= 10^{-8}$. The BBO  already proposed to achieve such sensitivity in measuring the chirp mass \citep{Crowder:2005nr} with a noise sensitivity of about $10^{-24}$ 1$/\sqrt{\rm Hz}$. A similar sensitivity is also possible from DECIGO, aTIANGO, and GLOC. A future experiment similar to these concepts, with a control on the detector noise to operate in the proposed sensitivity of these experiments, will be important.  {It is important to note that with a detector sensitivity of BBO (with an uncertainty on chirp mass $\sigma_{\mathcal{M}}/\mathcal{M}= 10^{-8}$), we can achieve only about a 5$\sigma$ detection of $w_0=-1$, but $w_a=0$ remains not so well constrained.} However, an improvement in the sensitivity to $10^{-24}$ 1$/\sqrt{\rm Hz}$ and laser frequency noise below $10^{-9}$ Hz$/\sqrt{\rm Hz}$ with future detector concept will be able to make a percent-level accurate measurement of $w_0$ and $w_a$ \citep{Cahillane:2021jvt},  {and can accurately capture any redshift evolution in the EoS of dark energy.} 
In summary, future BBO-like GW detectors will provide exquisite precision to understand dark energy, if multi-band observations are made. {However, the potential gain from this technique can be explored by considering next-generation detectors which are an order of magnitude better in sensitivity than BBO. Such detectors will make it possible to detect the evidence of dark energy without using any additional astrophysical distance or redshift calibration.}   

\section{Discussion and Conclusion}\label{sec5}
In this work, we show that in the framework of GR, observations of the shift in the detector frame chirp mass of the GW sources in two different frequency bands and inference of the luminosity distance provide a new way to measure the expansion rate of the Universe without requiring the use of any EM observations nor leveraging on any astrophysical assumption of the GW source.  Due to the advantage of the GW signal not being susceptible to foreground contamination such as dust, the measurement is possible up to high redshift, to the epoch of the first ever binary black hole in the Universe. 

This new frontier of observational cosmology, if realized from the GW detectors, can provide the best possible way to measure the high redshift Universe. However, such measurements will require control of several sources of errors, both on the instrument side as well as on the waveform modeling side. This kind of measurement will require waveform accuracy in the inspiral and merger phase up to $10^{-11}$ to successfully infer cosmology from GW signals. The degeneracy between inclination angle and luminosity distance \citep{PhysRevD.57.4535}, and impact from weak lensing \citep{2010PhRvD..81l4046H} are dominant sources of uncertainties on the luminosity distance side. On the chirp mass inference side, the chirp mass inference can be impacted by environmental effects. As we are only focusing on stellar origin BBHs, these black holes are unlikely to have accretion discs or be embedded only near a galactic center. As a result, the dephasing of the GW emission due to any local astrophysical effects is negligible. 
 
As we are at an exciting time of designing the next generation of GW detectors, appropriate planning for observational strategies and advanced detector sensitivities can make a paradigm shift in observational cosmology using this technique. This technique brings a way to measure the nature of dark energy, which is not limited by astrophysical modeling uncertainties and can be achieved by precise measurement from GW detectors operational from mili-Hertz to hecto-Hertz range. Beyond dark energy, this technique also makes it possible to independently map the expansion rate at high redshift, which can discover new components in the Universe that may be dominant only at high redshift. In summary, GW detector development, which can make robust and precise measurements of the chirp mass, can open a new window to fundamental physics. This technique makes GW a self-sufficient probe to explore cosmology without depending on any EM observations.  

\section*{Data Availability}
The data underlying this article will be shared at the request to the corresponding author.

\section*{Acknowledgements}
This work is part of the \texttt{⟨data|theory⟩ Universe-Lab}, funded by TIFR and the Department of Atomic Energy, Government of India. The author acknowledges the use of computer cluster of the \texttt{⟨data|theory⟩ Universe-Lab} and using Astropy \citep{2013A&A...558A..33A, 2018AJ....156..123A}, Giant-Triangle-Confusogram \citep{Bocquet2016}, IPython \citep{PER-GRA:2007}, Matplotlib \citep{Hunter:2007}, NumPy \citep{2011CSE....13b..22V}, and SciPy \citep{scipy} in this analysis.   
\bibliographystyle{mnras}
\bibliography{main_mnras}

\begin{thebibliography}{}
\makeatletter
\relax
\def\mn@urlcharsother{\let\do\@makeother \do\$\do\&\do\#\do\^\do\_\do\%\do\~}
\def\mn@doi{\begingroup\mn@urlcharsother \@ifnextchar [ {\mn@doi@} {\mn@doi@[]}}
\def\mn@doi@[#1]#2{\def\@tempa{#1}\ifx\@tempa\@empty \href {http://dx.doi.org/#2} {doi:#2}\else \href {http://dx.doi.org/#2} {#1}\fi \endgroup}
\def\mn@eprint#1#2{\mn@eprint@#1:#2::\@nil}
\def\mn@eprint@arXiv#1{\href {http://arxiv.org/abs/#1} {{\tt arXiv:#1}}}
\def\mn@eprint@dblp#1{\href {http://dblp.uni-trier.de/rec/bibtex/#1.xml} {dblp:#1}}
\def\mn@eprint@#1:#2:#3:#4\@nil{\def\@tempa {#1}\def\@tempb {#2}\def\@tempc {#3}\ifx \@tempc \@empty \let \@tempc \@tempb \let \@tempb \@tempa \fi \ifx \@tempb \@empty \def\@tempb {arXiv}\fi \@ifundefined {mn@eprint@\@tempb}{\@tempb:\@tempc}{\expandafter \expandafter \csname mn@eprint@\@tempb\endcsname \expandafter{\@tempc}}}

\bibitem[\protect\citeauthoryear{Aasi et~al.}{Aasi et~al.}{2015}]{LIGOScientific:2014pky}
Aasi J.,  et~al., 2015, \mn@doi [Class. Quant. Grav.] {10.1088/0264-9381/32/7/074001}, 32, 074001

\bibitem[\protect\citeauthoryear{Abbott et~al.}{Abbott et~al.}{2016}]{KAGRA:2013rdx}
Abbott B.~P.,  et~al., 2016, \mn@doi [Living Rev. Rel.] {10.1007/s41114-020-00026-9}, 19, 1

\bibitem[\protect\citeauthoryear{Abbott et~al.}{Abbott et~al.}{2023}]{KAGRA:2021duu}
Abbott R.,  et~al., 2023, \mn@doi [Phys. Rev. X] {10.1103/PhysRevX.13.011048}, 13, 011048

\bibitem[\protect\citeauthoryear{Acernese et~al.}{Acernese et~al.}{2015}]{VIRGO:2014yos}
Acernese F.,  et~al., 2015, \mn@doi [Class. Quant. Grav.] {10.1088/0264-9381/32/2/024001}, 32, 024001

\bibitem[\protect\citeauthoryear{Ade et~al.}{Ade et~al.}{2014}]{Planck:2013pxb}
Ade P. A.~R.,  et~al., 2014, \mn@doi [Astron. Astrophys.] {10.1051/0004-6361/201321591}, 571, A16

\bibitem[\protect\citeauthoryear{Aghanim et~al.}{Aghanim et~al.}{2020}]{Planck:2018vyg}
Aghanim N.,  et~al., 2020, \mn@doi [Astron. Astrophys.] {10.1051/0004-6361/201833910}, 641, A6

\bibitem[\protect\citeauthoryear{Ajith et~al.}{Ajith et~al.}{2025}]{Ajith:2024mie}
Ajith P.,  et~al., 2025, \mn@doi [JCAP] {10.1088/1475-7516/2025/01/108}, 01, 108

\bibitem[\protect\citeauthoryear{Akutsu et~al.}{Akutsu et~al.}{2021}]{KAGRA:2020tym}
Akutsu T.,  et~al., 2021, \mn@doi [PTEP] {10.1093/ptep/ptaa125}, 2021, 05A101

\bibitem[\protect\citeauthoryear{Ali-Ha\"\i{}moud, Kovetz  \& Kamionkowski}{Ali-Ha\"\i{}moud et~al.}{2017}]{Ali-Haimoud:2017rtz}
Ali-Ha\"\i{}moud Y.,  Kovetz E.~D.,   Kamionkowski M.,  2017, \mn@doi [Phys. Rev. D] {10.1103/PhysRevD.96.123523}, 96, 123523

\bibitem[\protect\citeauthoryear{{Amaro-Seoane} et~al.,}{{Amaro-Seoane} et~al.}{2017}]{2017arXiv170200786A}
{Amaro-Seoane} P.,  et~al., 2017, \mn@doi [arXiv e-prints] {10.48550/arXiv.1702.00786}, \href {https://ui.adsabs.harvard.edu/abs/2017arXiv170200786A} {p. arXiv:1702.00786}

\bibitem[\protect\citeauthoryear{{Astropy Collaboration} et~al.,}{{Astropy Collaboration} et~al.}{2013}]{2013A&A...558A..33A}
{Astropy Collaboration} et~al., 2013, \mn@doi [\aap] {10.1051/0004-6361/201322068}, \href {https://ui.adsabs.harvard.edu/abs/2013A&A...558A..33A} {558, A33}

\bibitem[\protect\citeauthoryear{{Astropy Collaboration} et~al.,}{{Astropy Collaboration} et~al.}{2018}]{2018AJ....156..123A}
{Astropy Collaboration} et~al., 2018, \mn@doi [\aj] {10.3847/1538-3881/aabc4f}, \href {https://ui.adsabs.harvard.edu/abs/2018AJ....156..123A} {156, 123}

\bibitem[\protect\citeauthoryear{{Binney} \& {Tremaine}}{{Binney} \& {Tremaine}}{1987}]{1987gady.book.....B}
{Binney} J.,  {Tremaine} S.,  1987, {Galactic dynamics}

\bibitem[\protect\citeauthoryear{Bocquet \& Carter}{Bocquet \& Carter}{2016}]{Bocquet2016}
Bocquet S.,  Carter F.~W.,  2016, \mn@doi [The Journal of Open Source Software] {10.21105/joss.00046}, 1

\bibitem[\protect\citeauthoryear{Cahillane, Mansell  \& Sigg}{Cahillane et~al.}{2021}]{Cahillane:2021jvt}
Cahillane C.,  Mansell G.,   Sigg D.,  2021, \mn@doi [Opt. Express] {10.1364/OE.439253}, 29, 42144

\bibitem[\protect\citeauthoryear{Chevallier \& Polarski}{Chevallier \& Polarski}{2001}]{Chevallier:2000qy}
Chevallier M.,  Polarski D.,  2001, \mn@doi [Int. J. Mod. Phys. D] {10.1142/S0218271801000822}, 10, 213

\bibitem[\protect\citeauthoryear{Cram{\'e}r}{Cram{\'e}r}{1946}]{1946c}
Cram{\'e}r H.,  1946, Scandinavian Actuarial Journal, 1946, 85

\bibitem[\protect\citeauthoryear{Crowder \& Cornish}{Crowder \& Cornish}{2005}]{Crowder:2005nr}
Crowder J.,  Cornish N.~J.,  2005, \mn@doi [Phys. Rev. D] {10.1103/PhysRevD.72.083005}, 72, 083005

\bibitem[\protect\citeauthoryear{{DESI Collaboration} et~al.,}{{DESI Collaboration} et~al.}{2024}]{2024arXiv240403002D}
{DESI Collaboration} et~al., 2024, \mn@doi [arXiv e-prints] {10.48550/arXiv.2404.03002}, \href {https://ui.adsabs.harvard.edu/abs/2024arXiv240403002D} {p. arXiv:2404.03002}

\bibitem[\protect\citeauthoryear{Dhani, Radice, Sch\"utte-Engel, Gardner, Sathyaprakash, Logoteta, Perego  \& Kashyap}{Dhani et~al.}{2024}]{Dhani:2023ijt}
Dhani A.,  Radice D.,  Sch\"utte-Engel J.,  Gardner S.,  Sathyaprakash B.,  Logoteta D.,  Perego A.,   Kashyap R.,  2024, \mn@doi [Phys. Rev. D] {10.1103/PhysRevD.109.044071}, 109, 044071

\bibitem[\protect\citeauthoryear{Ellis \& Vaskonen}{Ellis \& Vaskonen}{2020}]{Ellis:2020lxl}
Ellis J.,  Vaskonen V.,  2020, \mn@doi [Phys. Rev. D] {10.1103/PhysRevD.101.124013}, 101, 124013

\bibitem[\protect\citeauthoryear{Flanagan \& Hughes}{Flanagan \& Hughes}{1998}]{PhysRevD.57.4535}
Flanagan E.~E.,  Hughes S.~A.,  1998, \mn@doi [Phys. Rev. D] {10.1103/PhysRevD.57.4535}, 57, 4535

\bibitem[\protect\citeauthoryear{Hall et~al.}{Hall et~al.}{2021}]{Hall:2020dps}
Hall E.~D.,  et~al., 2021, \mn@doi [Phys. Rev. D] {10.1103/PhysRevD.103.122004}, 103, 122004

\bibitem[\protect\citeauthoryear{{Hawking} \& {Israel}}{{Hawking} \& {Israel}}{1989}]{1989thyg.book.....H}
{Hawking} S.~W.,  {Israel} W.,  1989, {Three Hundred Years of Gravitation}

\bibitem[\protect\citeauthoryear{{Hirata}, {Holz}  \& {Cutler}}{{Hirata} et~al.}{2010}]{2010PhRvD..81l4046H}
{Hirata} C.~M.,  {Holz} D.~E.,   {Cutler} C.,  2010, \mn@doi [\prd] {10.1103/PhysRevD.81.124046}, \href {https://ui.adsabs.harvard.edu/abs/2010PhRvD..81l4046H} {81, 124046}

\bibitem[\protect\citeauthoryear{Hunter}{Hunter}{2007}]{Hunter:2007}
Hunter J.~D.,  2007, \mn@doi [Computing In Science \& Engineering] {10.1109/MCSE.2007.55}, 9, 90

\bibitem[\protect\citeauthoryear{Inayoshi, Tamanini, Caprini  \& Haiman}{Inayoshi et~al.}{2017}]{PhysRevD.96.063014}
Inayoshi K.,  Tamanini N.,  Caprini C.,   Haiman Z.,  2017, \mn@doi [Phys. Rev. D] {10.1103/PhysRevD.96.063014}, 96, 063014

\bibitem[\protect\citeauthoryear{Jani \& Loeb}{Jani \& Loeb}{2020}]{Jani:2020gnz}
Jani K.,  Loeb A.,  2020, \mn@doi [arXiv: 2007.08550] {10.1088/1475-7516/2021/06/044}

\bibitem[\protect\citeauthoryear{Jones, Oliphant, Peterson  et~al.}{Jones et~al.}{01  }]{scipy}
Jones E.,  Oliphant T.,  Peterson P.,   et~al., 2001--, {SciPy}: Open source scientific tools for {Python}, \url {http://www.scipy.org/}

\bibitem[\protect\citeauthoryear{Kawamura et~al.}{Kawamura et~al.}{2006}]{Kawamura:2006up}
Kawamura S.,  et~al., 2006, \mn@doi [Class. Quant. Grav.] {10.1088/0264-9381/23/8/S17}, 23, S125

\bibitem[\protect\citeauthoryear{Kuns, Yu, Chen  \& Adhikari}{Kuns et~al.}{2020}]{Kuns:2019upi}
Kuns K.~A.,  Yu H.,  Chen Y.,   Adhikari R.~X.,  2020, \mn@doi [Phys. Rev. D] {10.1103/PhysRevD.102.043001}, 102, 043001

\bibitem[\protect\citeauthoryear{Linder}{Linder}{2003}]{PhysRevLett.90.091301}
Linder E.~V.,  2003, \mn@doi [Phys. Rev. Lett.] {10.1103/PhysRevLett.90.091301}, 90, 091301

\bibitem[\protect\citeauthoryear{{Loeb}}{{Loeb}}{1998}]{1998ApJ...499L.111L}
{Loeb} A.,  1998, \mn@doi [\apjl] {10.1086/311375}, \href {https://ui.adsabs.harvard.edu/abs/1998ApJ...499L.111L} {499, L111}

\bibitem[\protect\citeauthoryear{Madau \& Dickinson}{Madau \& Dickinson}{2014}]{Madau:2014bja}
Madau P.,  Dickinson M.,  2014, \mn@doi [Ann. Rev. Astron. Astrophys.] {10.1146/annurev-astro-081811-125615}, 52, 415

\bibitem[\protect\citeauthoryear{Maggiore et~al.}{Maggiore et~al.}{2020}]{Maggiore:2019uih}
Maggiore M.,  et~al., 2020, \mn@doi [JCAP] {10.1088/1475-7516/2020/03/050}, 03, 050

\bibitem[\protect\citeauthoryear{P\'erez \& Granger}{P\'erez \& Granger}{2007}]{PER-GRA:2007}
P\'erez F.,  Granger B.~E.,  2007, \mn@doi [Computing in Science and Engineering] {10.1109/MCSE.2007.53}, 9, 21

\bibitem[\protect\citeauthoryear{{Perlmutter} et~al.,}{{Perlmutter} et~al.}{1997}]{1997ApJ...483..565P}
{Perlmutter} S.,  et~al., 1997, \mn@doi [\apj] {10.1086/304265}, \href {https://ui.adsabs.harvard.edu/abs/1997ApJ...483..565P} {483, 565}

\bibitem[\protect\citeauthoryear{Poisson \& Will}{Poisson \& Will}{1995}]{Poisson:1995ef}
Poisson E.,  Will C.~M.,  1995, \mn@doi [Phys. Rev. D] {10.1103/PhysRevD.52.848}, 52, 848

\bibitem[\protect\citeauthoryear{{Punturo} et~al.,}{{Punturo} et~al.}{2010}]{2010CQGra..27h4007P}
{Punturo} M.,  et~al., 2010, \mn@doi [Classical and Quantum Gravity] {10.1088/0264-9381/27/8/084007}, \href {https://ui.adsabs.harvard.edu/abs/2010CQGra..27h4007P} {27, 084007}

\bibitem[\protect\citeauthoryear{Raidal, Vaskonen  \& Veerm\"ae}{Raidal et~al.}{2017}]{Raidal:2017mfl}
Raidal M.,  Vaskonen V.,   Veerm\"ae H.,  2017, \mn@doi [JCAP] {10.1088/1475-7516/2017/09/037}, 09, 037

\bibitem[\protect\citeauthoryear{Rao}{Rao}{1945}]{1945r}
Rao C.~R.,  1945, Reson. J. Sci. Educ, 20, 78

\bibitem[\protect\citeauthoryear{{Riess} et~al.,}{{Riess} et~al.}{1998}]{1998AJ....116.1009R}
{Riess} A.~G.,  et~al., 1998, \mn@doi [\aj] {10.1086/300499}, \href {https://ui.adsabs.harvard.edu/abs/1998AJ....116.1009R} {116, 1009}

\bibitem[\protect\citeauthoryear{Ronchini et~al.,}{Ronchini et~al.}{2022}]{Ronchini:2022gwk}
Ronchini S.,  et~al., 2022, \mn@doi [Astron. Astrophys.] {10.1051/0004-6361/202243705}, 665, A97

\bibitem[\protect\citeauthoryear{{Sandage}}{{Sandage}}{1962}]{1962ApJ...136..319S}
{Sandage} A.,  1962, \mn@doi [\apj] {10.1086/147385}, \href {https://ui.adsabs.harvard.edu/abs/1962ApJ...136..319S} {136, 319}

\bibitem[\protect\citeauthoryear{{Schmidt} et~al.,}{{Schmidt} et~al.}{1998}]{1998ApJ...507...46S}
{Schmidt} B.~P.,  et~al., 1998, \mn@doi [\apj] {10.1086/306308}, \href {https://ui.adsabs.harvard.edu/abs/1998ApJ...507...46S} {507, 46}

\bibitem[\protect\citeauthoryear{{Schutz}}{{Schutz}}{1986}]{1986NaturS}
{Schutz} B.~F.,  1986, \mn@doi [\nat] {10.1038/323310a0}, \href {https://ui.adsabs.harvard.edu/abs/1986Natur.323..310S} {323, 310}

\bibitem[\protect\citeauthoryear{Seto, Kawamura  \& Nakamura}{Seto et~al.}{2001}]{Seto:2001qf}
Seto N.,  Kawamura S.,   Nakamura T.,  2001, \mn@doi [Phys. Rev. Lett.] {10.1103/PhysRevLett.87.221103}, 87, 221103

\bibitem[\protect\citeauthoryear{Shlivko \& Steinhardt}{Shlivko \& Steinhardt}{2024}]{Shlivko:2024llw}
Shlivko D.,  Steinhardt P.,  2024, arXiv: 2405.03933

\bibitem[\protect\citeauthoryear{{Spergel} et~al.,}{{Spergel} et~al.}{2003}]{2003ApJS..148..175S}
{Spergel} D.~N.,  et~al., 2003, \mn@doi [\apjs] {10.1086/377226}, \href {https://ui.adsabs.harvard.edu/abs/2003ApJS..148..175S} {148, 175}

\bibitem[\protect\citeauthoryear{Tegmark et~al.}{Tegmark et~al.}{2004}]{SDSS:2003eyi}
Tegmark M.,  et~al., 2004, \mn@doi [Phys. Rev. D] {10.1103/PhysRevD.69.103501}, 69, 103501

\bibitem[\protect\citeauthoryear{{van der Walt}, {Colbert}  \& {Varoquaux}}{{van der Walt} et~al.}{2011}]{2011CSE....13b..22V}
{van der Walt} S.,  {Colbert} S.~C.,   {Varoquaux} G.,  2011, \mn@doi [Computing in Science and Engineering] {10.1109/MCSE.2011.37}, \href {https://ui.adsabs.harvard.edu/abs/2011CSE....13b..22V} {13, 22}

\makeatother
\end{thebibliography}
\label{lastpage}
\end{document}